\documentclass[iop,apj]{emulateapj}
\usepackage{amsmath,amssymb,amstext}

\usepackage[breaklinks,colorlinks,citecolor=blue,linkcolor=magenta]{hyperref} 

\usepackage[all]{hypcap}
\usepackage{natbib}
\usepackage{epstopdf}
\usepackage{rotating}
\usepackage{txfonts}

\shorttitle{White-light continuum of superflares}
\shortauthors{Heinzel \& Shibata}

\begin{document}

\title{Can flare loops contribute to the white-light emission of stellar superflares ?}
\author{P. Heinzel\altaffilmark{1,2} \& K. Shibata\altaffilmark{2}}

\email{pheinzel@asu.cas.cz}

\altaffiltext{1}{Astronomical Institute, Czech Academy of Sciences, 25165 Ond\v{r}ejov, Czech Republic}
\altaffiltext{2}{Graduate School of Science, Kyoto University, Kyoto 606-8502, Japan}  

\begin{abstract}
Since the discovery of stellar superflares by $Kepler$ satellite, these extremely energetic events have been studied in 
analogy to solar flares. Their white-light (WL) continuum emission has been interpreted as being produced by heated 
ribbons. In this paper we compute the WL emission from overlying flare loops depending on their density and temperature and show
that, under conditions expected during superflares, the continuum brightening due to extended loop arcades can significantly contribute to stellar flux detected
by $Kepler$. This requires electron densities in the loops $10^{12} - 10^{13}$ cm$^{-3}$ or higher. We show that such densities, exceeding those
typically present in solar flare loops, can be reached on M-dwarf and solar-type superflare stars with large starspots and much stronger magnetic fields.
Quite importantly, the WL radiation of loops is not very sensitive to their temperature and thus both cool as well as hot loops may contribute.
We show that the WL intensity emergent from optically-thin loops is lower than the blackbody radiation from flare ribbons, 
but the contribution of loops to total stellar flux can be quite important due to their significant emitting areas.
This new scenario for interpreting superflare emission suggests that the observed WL flux is due to a mixture of the ribbon and loop radiation
and can be even loop-dominated during the gradual phase of superflares.
\end{abstract}

\maketitle

\section{Introduction}

After the recent discovery of stellar superflares in $Kepler$ satellite data (\cite{Maehara2012}), a series
of papers rapidly emerged trying to explain this phenomenon mainly in terms of correlations 
between various stellar parameters. The total flare energy was estimated from the
light curves, for statistical samples
of both dMe red-dwarf and G-type solar-like stars, and was related to flare duration, magnetic activity (represented by large starspots), stellar rotation velocity, stellar age, etc.
(\cite{Notsu+2015}, \cite{Shibata2016}, \cite{Shulyak+2017}, \cite{Namekata2017}).
However, only a limited attention was devoted to understanding the mechanisms of the superflare emission.
First of all, $Kepler$ light curves represent the white-light (WL) emission of a flare, integrated
in the broad optical passband from 400 to 900 nm. Therefore, all $Kepler$ superflares are, according to
standard solar terminology, the so-called White Light Flares (WLF). But superflares on
cool stars seem to have much larger total power than solar flares, i.e. up to 10$^{38}$ erg compared
with the maximum of $10^{32}$ erg in the solar cases (\cite{Shibata2016}). This follows from the time integrated fluxes measured by {\em Kepler}.

A plausible scenario of the energy release in superflares is the same as for solar flares, i.e. the magnetic reconnection in the corona and the energy transport down to low atmospheric layers by 
particle beams (mainly electrons) and by thermal conduction. This then suggests that on stars we observe the surface structures analogical
to solar {\em flare ribbons} where all solar WLFs are normally detected. On the Sun the WL
ribbons usually disappear after the flare impulsive phase, while the decaying ribbons continue to be visible in various spectral lines during the {\em gradual phase}. This later phase is also characterised by typical appearance of the so called 'post-flare loops', visible in different spectral lines. 
However, to our
knowledge these loops have never been detected in the white light against the solar 
disk and only recently, quite importantly, they have been rarely seen high above the limb during the off-limb
observations of strong flares by {\em SDO}/HMI (\cite{S-Hilaire+2014}, \cite{Krucker2015}). Therefore, on the Sun these
loops don't affect the WLF emission. If the same applies also for stellar superflares,
we can think that their strong WL emission emerges entirely from ribbons and this is how the energy released during the reconnection is normally related to WL emission. 
However, with an increasing spatial resolution of
solar-flare observations, we find that the solar ribbons are actually very narrow (although often long)
features, occupying only a tiny fraction of the active region area - see the latest GST movie in \cite{Jing2016}, 
while the cool H$\alpha$ loops cover much large space.

Extrapolating this to superflares where the magnetic loops are expected to be much larger and assuming that the WL visibility of
flare loops during superflares can be quite different compared to the solar case, we investigate in this study whether such loops 
can contribute to WL emission of superflares. Based on our theoretical analysis, we propose that the flare loops can indeed
significantly contribute to total superflare WL emission, if not dominate it totally, because
they can be visible against the stellar disk (namely in case of very cool dMe stars) and they
can occupy much larger areas than just the ribbons. This may then
change our current picture of the WL emission of stellar superflares, but also the generally accepted
paradigm of stellar-flare optical emission as being entirely due to flare ribbons. The radiation-hydrodynamical (RHD) models of stellar
flares are one-dimansional models of a flare loop where all emissions in cooler lines and continua arise from the
ribbons (\cite{Kowalski2016}). These models didn't predict yet the optical emission from overlying loops which are
typically observed after an impulsive onset of solar flares. In particular the line emission
must be even stronger in case of superflares and, moreover, we predict here the importance of WL emission of such loops.

The paper is organised as follows. In Section 2 we describe the loops typically observed during
solar flares and discuss their visibility. Section 3 summarizes all relevant mechanisms of the WL emission 
produced by flare loops. In Section 4 we present our numerical results of the WL visibility of loops during stellar superflares.
Since the critical parameter of such a visibility is the electron density in the loops, we
make density estimates for the case of superflares in Section 5 using various physical arguments. 
In Section 6 we comment on spectral characteristics of stellar WL flares and 
Section 7 contains discussion and conclusions.

\section{Visibility of loops during solar flares}

In the majority of solar flares, including WLFs, we see dark H$\alpha$ loops
during the gradual phase - see e.g. an excellent movie of the X1 flare
taken at very high resolution with GST (\cite{Jing2016}). 
These cool loops have been, somewhat misleadingly, called 'post-flare' loops (\cite{Svestka2007}), in this paper
we will call them simply flare loops.
To see cool flare loops against the solar disk in absorption in the H$\alpha$ line, the electron density in the loop
should not exceed $10^{12}$ cm$^{-3}$ (\cite{HK1987}). The same loops, however,
can be seen in emission in other lines like MgII h \& k (\cite{Lacatus+2017}, \cite{Mikula+2017}). At such
densities or lower, the loops will not be visible in WL against the disk, but can
be detected as faint WL loops high above the limb (\cite{S-Hilaire+2014}, 
\cite{Krucker2015}). Those loops,
however, should not be confused with the off-limb WL flare emission at chromospheric 
heights (\cite{BK2011}, \cite{Krucker2015}, \cite{Heinzel+2017}). High-lying
WL flare loops emit either in the Paschen recombination continuum with free-free
contribution (depending on the temperature), or due to Thomson scattering of the photospheric radiation on
loop electrons. For electron densities lower than $10^{12}$ cm$^{-3}$, the latter
mechanism should dominate (\cite{Heinzel+2017}). 
Cool loops form from initially hot ones (10$^7$ K) which have been formed by the evaporation 
during strong chromospheric and transition-region
heating. Depending on the densities, the loops at different temperatures appear within typical
cooling times which depend on the conductive and radiative cooling processes. In the solar case, for electron densities of cool loops
typically lower than $10^{12}$ cm$^{-3}$, the cooling time roughly takes minutes and this may explain why we see the cool loops later after the
impulsive heating with electron beams is already over (HXR is also no longer detectable). However, in case of superflares,
where we expect larger densities, cooler loops may appear practically immediately after the flare onset.

\section{Mechanisms of the white-light continuum emission}

In the wavelength range of {\em Kepler}, the WL continuum emitted by dense superflare loops will be mainly due
to the hydrogen recombination (namely the Paschen continuum) and due to the hydrogen free-free process. 
Below we detail these mechanisms, for general formulae see also \cite{HM2015}).

\subsection{Hydrogen recombination continua}

At low temperatures, the loops are partially ionized and assuming their temperature $T$ and electron density $n_e$ as free parameters, 
we can compute the WL radiation. The absorption coefficient for bound-free hydrogen transitions from atomic level $i$ is

\begin{equation}
\kappa_{\nu}^{\rm bf} = \alpha_{\nu} (n_i - n_i^{*} e^{-h \nu / k T}) \, ,
\end{equation}
where $\alpha_{\nu}$ is the hydrogen photoionization cross-section

\begin{equation}
\alpha_{\nu} = 2.815 \times 10^{29} g_{\rm bf}(i,\nu) / i^5 / \nu^3
\end{equation} 
with $g_{\rm bf}$ being the Gaunt factor for bound-free opacity. $n_i$ is the non-LTE population of the hydrogen level $i$ from which the
photoionization takes place and $n_i^*$ is its LTE counterpart. The second term represents the stimulated emission which is normally
treated as a negative absorption in the radiative-transfer equation. 
The departure coefficient from LTE is defined as $b_i = n_i/n_i^*$.
$h$ and $k$ are the Planck and Boltzmann constants, respectively. Since we are not solving here the full non-LTE radiative-transfer problem,
we have to make some assumption about $b_i$. At relatively high densities which we will consider for the flare loops, detailed non-LTE
modeling shows that the $b$-factors for third and higher hydrogen levels are close to unity. We thus
assume here that $b_3$ (Paschen continuum) and $b_4$ (Brackett continuum) are equal to one; we neglect the opacity of higher continua
in the WL range observed by {\em Kepler}. However, we will show later that even departures from unity of these factors don't affect our results
substantially. With $b_i = 1$ we thus get

\begin{equation}
\kappa_{\nu}^{\rm bf} = \alpha_{\nu} n_i^{*}(1 - e^{-h \nu / k T}) \, ,
\end{equation}
with 
\begin{equation}
n_i^* = n_p n_e \Phi_i (T) \, ,
\end{equation}
where $\Phi_i (T) = 2.0707 \times 10^{-16} 2i^2 e^{h\nu_i/kT}/T^{3/2}$ is the Boltzmann factor.
In what follows we assume a pure hydrogen plasma for which
$n_p n_e=n_e^2$.
Finally, it can be shown that the source function of the Paschen and Brackett continuum is approximately equal to

\begin{equation}
S_{\nu} \simeq \frac{1}{b_i} B_{\nu}(T) \simeq B_{\nu}(T) \, .
\end{equation}

\subsection{Hydrogen free-free continuum}

Hydrogen free-free opacity is expressed as

\begin{equation}
\kappa_{\nu}^{\rm ff} = 3.69 \times 10^8 n_p n_e g_{\rm ff}(\nu, T) T^{-1/2} \nu^{-3} (1 - e^{-h \nu / k T}) \, ,
\end{equation}
with the Gaunt factor $g_{\rm ff}$.
The source function is equal to Planck function for the free-free process.

\subsection{Radiative transfer}

Assuming a loop of diameter $D$, having a uniform continuum source function $S_{\nu}$ and observed against the stellar disk, 
we can express the emergent loop intensity as the formal solution of radiative-transfer equation

\begin{equation}
I_{\nu} = I_{\rm bg}  e^{-\tau_{\nu}} + S_{\nu} (1 - e^{-\tau_{\nu}})  \, ,
\end{equation}
where the first term represents the background radiation from the stellar disk below the loop $I_{\rm bg}$, attenuated by the continuum loop opacity
with $\tau_{\nu}$ being the optical thickness of the loop at a given continuum frequency. The second term is the radiation intensity of the loop itself. 
Note that for optically-thin loops, the resulting intensity is a mixture of partially penetrating background intensity and the
radiation intensity of the loop itself.
Using the approximation $b_i=1$, we set $S_{\nu}=B_{\nu}(T)$. The total optical thickness is 

\begin{equation}
\tau_{\nu} = (\kappa_{\nu}^{\rm bf} + \kappa_{\nu}^{\rm ff}) D \, .
\end{equation}

\subsection{Thomson scattering on loop electrons}

The Thomson scattering is process different from the free-bound and free-free emission and we include it here as an
optically thin emission, simply added to the intensity as computed by Eq. (7). More precise radiative-transfer calculations are
not needed because (i) fb and ff opacity will be shown to be small as well, and (ii) the Thomson scattering is practically negligible
for most conditions considered in this study. Contribution to the loop intensity is expressed as

\begin{equation}
I_{\nu}^{\rm Th} = n_e \sigma_{\rm T} J_{\nu}^{\rm inc} D\, ,
\end{equation}
where $\sigma_{\rm T} = 6.65 \times 10^{-25}$ cm$^2$  is the cross-section for Thomson scattering and the incident intensity $J_{\nu}^{\rm inc}$ is computed as the product
of $B_{\nu}(T_{\rm eff})$ and geometrical dilution factor. $T_{\rm eff}$ is the effective temperature of the star under consideration.
For geometrical dilution factor we take a value 0.4 which corresponds to loop heights around 10000 km.
Note that both fb and ff emissivities are proportional to $n_e^2$ while the Thomson scattering scales only linearly
with $n_e$. 
For M dwarfs the incident radiation is much weaker than that for solar-type stars (low $T_{\rm eff}$) and it will be even lower above large areas occupied
by dark starspots.
    
\section{Computed flux amplitudes due to flare loops}

\subsection{Flare areas}

The flux of the entire preflare star is

\begin{equation}
F_s = (A_s - A_f) I_s + A_f I_{\rm bg} \, ,
\end{equation}
where $A_s=\pi R^2$ is the stellar surface ($R$ being the stellar radius), $A_f$ is the flare area, and $I_s$ represents the stellar surface intensity computed
from the effective temperature $T_{\rm eff}$ of the star.
During a flare the stellar flux will be 

\begin{equation}
F = A_f I + (A_s - A_f) I_s \, .
\end{equation}
The $Kepler$ flux amplitude relative to the preflare flux is then defined as

\begin{equation}
\frac{\Delta F}{F_s} = \frac{F - F_s}{F_s} = \frac{A_f}{A_s} \frac{I - I_{\rm bg}}{I_s-\frac{A_f}{A_s} (I_s - I_{\rm bg})} \, .
\end{equation}
We consider two limiting cases for $I_{\rm bg}$, $I_{\rm bg}=I_s$ (i.e. no starspots) and $I_{\rm bg}=I_{\rm spot}$ meaning that the preflare
area is covered by starspots having the radiation temperature $T_{\rm spot}$. The former case is usually used in the literature as approximation (\cite{Shibayama+2013}),
in this study we assume that $I_{\rm bg} = I_{\rm spot}$. 
For simplicity of exposition we have omitted here
the frequency indexes, in the following we will show the results for a peak wavelength 600 nm of the {\em Kepler} WL passband.
For a more precise comparison with $Kepler$ observations, one has to integrate fluxes over the whole wavelength range of {\em Kepler}, 
weighted by its transmission profile. This is shown e.g. in \cite{Shibayama+2013}, together with the procedure of computing the total flare energy, 
integrated over flare lifetime.

\subsection{Theoretical amplitudes}

    \begin{figure}
    \centering
%    \figurenum{1}              
    \includegraphics[height=6.8cm, angle=0]{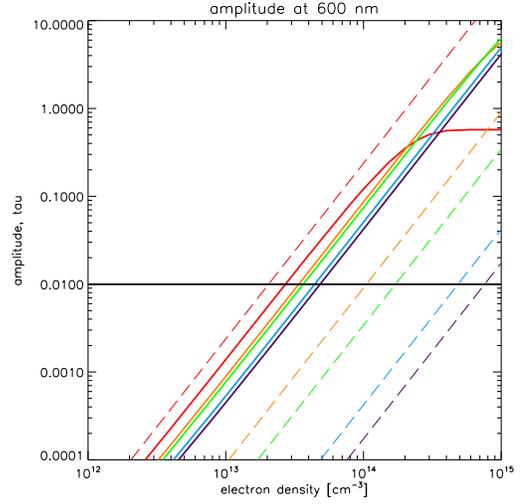}
    \caption{Theoretical amplitude (full lines) and optical thickness (dashed lines) 
    of the WL loop at 600 nm. Model with $T_{\rm eff}=6000$ K, $T_{\rm spot}=4000$ K, $D$=1000 km, $A_f/A_s=$0.1\, .
    Loop temperatures: 10000 K - red, 50000 K - orange, $10^5$ K - green, $5\times 10^5$ K - blue, 1 MK - magenta. Horizontal
    black line represents a characteristic amplitude from {\em Kepler} observations for G and M stars.}
    \label{fig1}
    \end{figure}
    \begin{figure}
    \centering
%    \figurenum{2}              
    \includegraphics[height=6.8cm, angle=0]{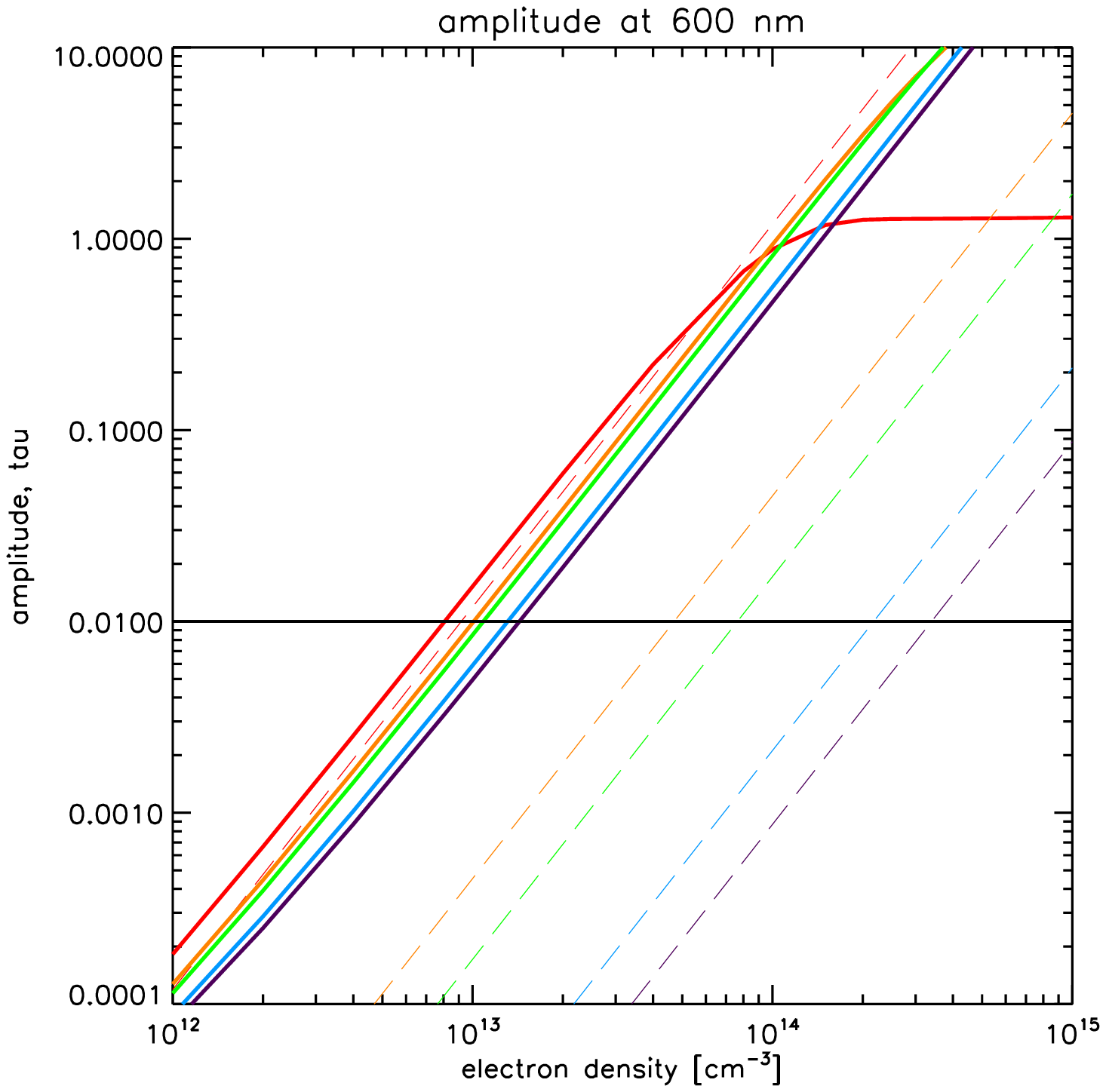}
    \caption{Same as in Fig. 1, but for model with $T_{\rm eff}=6000$ K, $T_{\rm spot}=4000$ K, $D$=5000 km, $A_f/A_s=$0.2 \, .}
    \label{fig2}
    \end{figure}
    \begin{figure}
    \centering
%    \figurenum{3}              
    \includegraphics[height=6.8cm, angle=0]{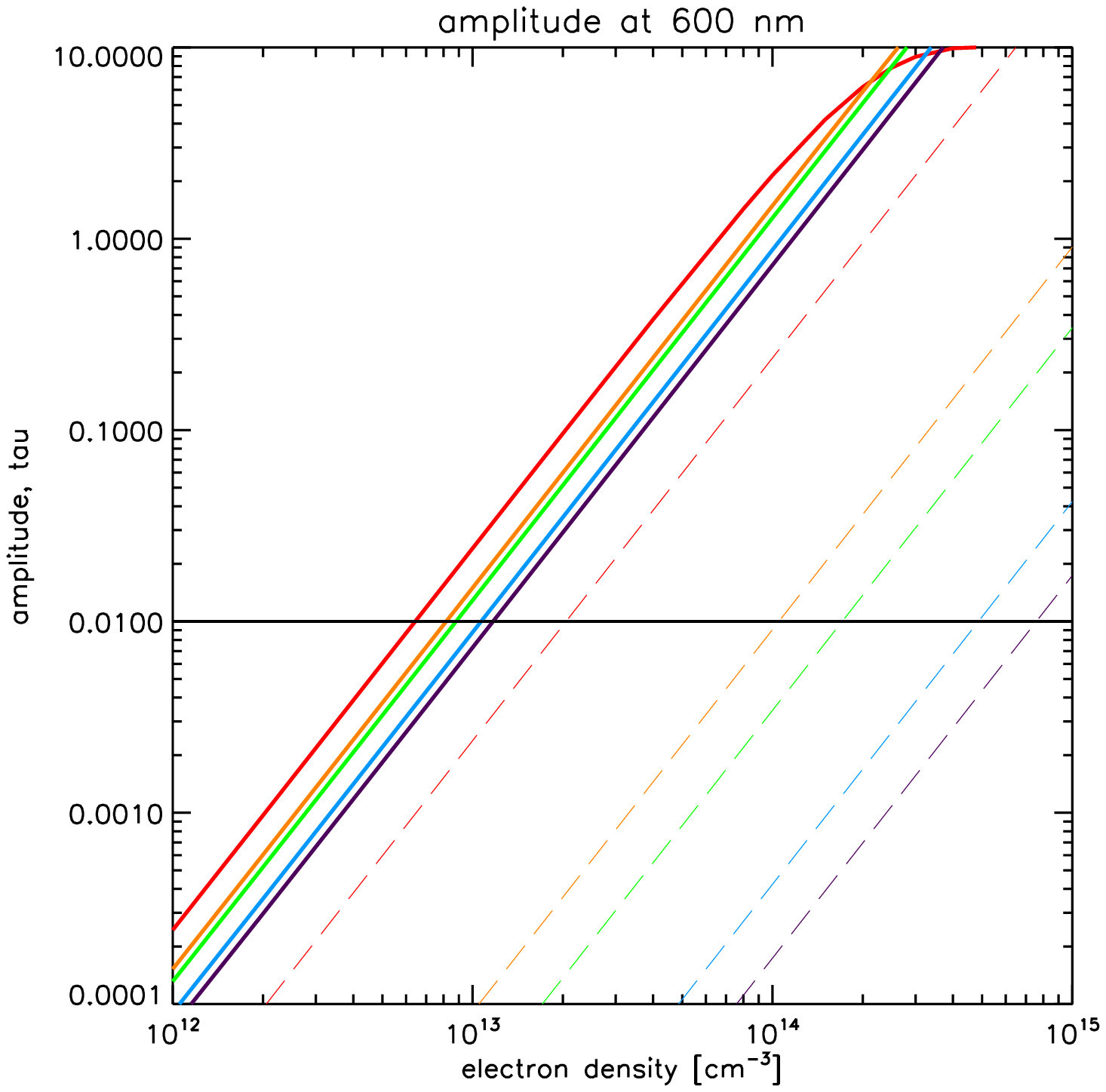}
    \caption{Same as in Fig. 1, but for model with $T_{\rm eff}=3500$ K, $T_{\rm spot}=3000$ K, $D$=1000 km, $A_f/A_s=$0.1 \, .}
    \label{fig3}
    \end{figure}
    \begin{figure}
    \centering
%    \figurenum{4}      
    \includegraphics[height=6.8cm, angle=0]{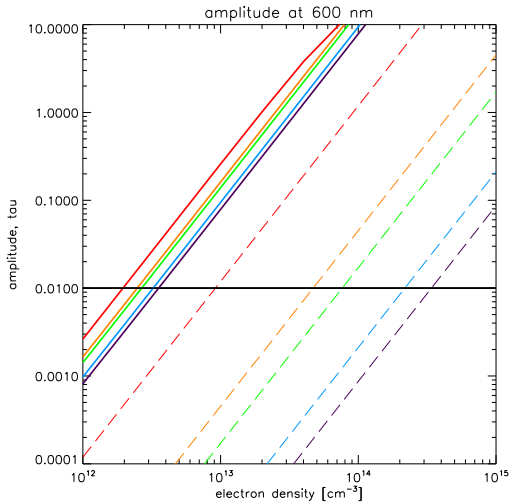}
    \caption{Same as in Fig. 1, but for model with $T_{\rm eff}=3500$ K, $T_{\rm spot}=3000$ K, $D$=5000 km, $A_f/A_s=$0.2.}
    \label{fig4}
    \end{figure}

We have computed the flare intensity $I$ for a variety of temperatures and electron densities expected in superflare loops, assuming
different stellar $T_{\rm eff}$, different $I_{\rm bg}$ and varying $D$. Then the flux amplitudes $\Delta F/F_s$ were evaluated using parametric values of 
the filling factor $A_f/A_s$.  
From various models we have selected four representative examples to show in this exploratory paper, more extensive analysis will be done in a next paper.
All figures 1 - 4 show variations of theoretical amplitudes with the electron density, for 5 temperatures ranging from 10000 K to 1 MK. In the same figures we also
plot the continuum optical thickness of the loops at 600 nm (dashed lines). In general the amplitudes show a linear increase with increasing electron density in our
log-log plots, which applies for the optically-thin regime. This means that the flare intensity is a sum of partially attenuated background radiation and of the emission by the loop
itself. However, when cool loops become optically thick at high electron densities (typically greater than 10$^{14}$ cm$^{-3}$), 
the intensity saturates to the blackbody value corresponding to kinetic temperature of the loop and is no longer dependent on $n_e$. This is well
visible in Figs. 1 and 2 for $T$=10000 K. Both optically-thin and saturated regimes also imply that the recombination component of the emergent radiation is
independent on the value of $b$-factors so that our assumption that $b_3=1$  and $b_4=1$ is justified in these two regimes. 
Figs. 1 and 2 show the situation for a solar-type star with $T_{\rm eff}$ = 6000 K,
with a spot covering the flare region ($T_{\rm spot}=4000$ K). Fig. 1 shows the case with $D$=1000 km, a single loop having typical diameter as known from
solar observations, and the filling factor 0.1 which is a rather conservative estimate for stellar active regions. Fig. 2 is for $D$=5000 km and a larger flare area having
the filling factor 0.2 .
In these two cases, we would need electron densities between 10$^{13}$ and 10$^{14}$ cm$^{-3}$ to get the amplitude consistent
with characteristic {\em Kepler} observations represented by the black horizontal line in all figures. 
However, the situation with M-type stars is different due to low luminosity of the stellar disk at $T_{\rm eff}$ = 3500 K and $T_{\rm spot}=3000$ K
(we estimate the spot temperatures according to \cite{Berdyugina2005}).
We immediately see that for cool M-type stars the flare loops can significantly contribute to WL amplitudes already at electron densities
somewhat higher than 10$^{12}$ cm$^{-3}$. As noticed in section 2, this density is a limiting one at which the H$\alpha$ loops are still visible in absorption against
the solar disk. Densities higher than 10$^{12}$ cm$^{-3}$ require stronger heating and evaporation process which we discuss in the next section. 

Another very important finding is that the range of electron densities needed to get the observed amplitude is rather narrow, for loop temperatures spanning  
two orders of magnitude. This means that both cool loops, as well as hot loops, may contribute significantly to the overall
WL emission. The contribution of hot loops is somewhat smaller so that we need a bit higher electron density to reach the observed amplitude. 
During solar flares we actually see cool and hot loops at the same time
(e.g. \cite{Mikula+2017}), while on stars a mixture of such loops will contribute to the total stellar flux.

\subsection{Relative importance of WL emissivities}

Since we solve the transfer equation for three overlapping continua (i.e. Paschen, Brackett and free-free), it is not directly evident which mechanism dominates
the emergent intensity and thus the amplitudes for given values of $T$ and $n_e$. However, we found that the most typical situation is that flare loops are optically thin for moderate
electron densities considered in our grid of models. In such a case, the ratio of optically-thin emissivities of the free-free and Paschen continuum takes the
simple form (\cite{Heinzel+2017}

\begin{equation}
\frac{I^{\rm ff}}{I^{\rm Pa}} = 8.55 \times 10^{-5} T  e^{-h\nu_i/kT}
\end{equation}
assuming that both Gaunt factors are unity (note that in \cite{Heinzel+2017} the exponent is positive by a misprint but the computed ratios are correct).
This ratio is independent of the electron density and is only the function of temperature. We see that in the optically-thin regime the free-free
emission starts to dominate the Paschen recombination at temperatures higher than about $3 \times 10^4$ K. At $T=10^4$ K the ratio is 0.15 but
at 1 MK the free-free emission dominates the free-bound ones completely. Therefore, at high temperatures the WL loop emission is dominated by free-free
process, especially at electron densities higher than 10$^{12}$ cm$^{-3}$. 

\section{Electron densities and cooling time scales}

From the previous section we can see under which conditions the flare loops can substantially contribute to the WL emission of superflares. In this section
we try to estimate the electron densities as a key parameter, by extrapolating our experience with solar flares to superflares. 

According to the standard magnetic-reconnection model of solar flares (e.g. \cite{ShiMag2011}), the energy flux $F$ (erg cm$^{-2}$ s$^{-1}$), 
injected into the footpoint of the reconnected loop, can be written as

\begin{equation} 
F = \frac{B^2}{4 \pi} V_A = 10^{12} {\rm erg} \, {\rm cm}^{-2} {\rm s}^{-1} \bigg(\frac{B}{100 \, {\rm G}} \bigg)^3 \bigg(\frac{n_e}{10^9 \, {\rm cm}^{-3}} \bigg)^{-1/2} \, ,
\end{equation}
where $B$ (G) is the magnetic field strength in the inflow region of the flare current sheet, $n_e$ (cm$^{-3}$) is the electron density in the inflow region, and 
$V_A$ is the Alfv\'{e}n speed in the inflow region (\cite{YokoShi1998}, \cite{YokoShi2001}).
Let's assume the fraction $a$ of this energy flux is used to accelerate high energy electrons. Then, the energy flux $F_e$ of nonthermal electron beams can be written
as $F_e = a F$. When this electron beam collides with the dense chromospheric plasma with density $n_{\rm ch}$, the beam energy is thermalized to heat the chromospheric plasma
from 10$^4$ K to flare temperatures $T$. 

The radiation cooling time $t_{\rm rad}$ (sec) for the chromospheric plasma with density $n_{\rm ch}$ (cm$^{-3}$) can be expressed as

\begin{equation} 
t_{\rm {rad}}  =  \frac{3n_{\rm ch} kT}{n_{\rm ch}^2 \Lambda(T)}
=  100 \, {\rm s} \, \bigg(\frac{n_{\rm ch}}{10^{12} \, {\rm cm}^{-3}} \bigg)^{-1} \bigg(\frac{T}{10^7 \, {\rm K}} \bigg)^{5/3} \, ,
\end{equation}
where we used $\Lambda(T)=10^{-17.73} T^{-2/3}$ for 6.3 $\le \log T \le$ 7.0
In this case, radiation cooling time is too long to balance with the heating by electron beams. Hence the heated chromospheric plasma suddenly expands upward leading to chromospheric evaporation and downward leading to shock and chromospheric condensation. 

Roughly speaking, the flare temperature is determined by the balance between reconnection heating and conduction cooling (\cite{YokoShi1998}, \cite{YokoShi2001}) even if the main energy transport from the reconnection site to the chromosphere is due to nonthermal electron beams. This is because the nonthermal electrons quickly thermalize in the upper chromosphere, and then the thermal conduction finally transports heat throughout the flare loop. 
Then the flare temperature $T$ (K) can be written (\cite{YokoShi1998}, \cite{YokoShi2001}, \cite{ShiYoko1999}, \cite{ShiYoko2002})

\begin{equation}
\frac{\kappa_0 T^{7/2}}{L} = F_e = a \frac{B^2}{4 \pi} V_A \, .
\end{equation}
Namely we get 

\begin{equation} 
T \simeq 3 \times 10^7 \, {\rm K} \, \bigg(\frac{F_e}{10^{11} \, {\rm erg} \, {\rm cm}^{-2} \, {\rm s}^{-1}} \bigg)^{2/7} \bigg(\frac{L}{10^9 \, {\rm cm}} \bigg)^{2/7} \, .
\end{equation}
Note that a reference electron-beam flux of 10$^{11}$ erg cm$^{-2}$ s$^{-1}$ is obtained for $B$=100 G, $n_e$=10$^9$ cm$^{-3}$ and $a$=0.1 .
In reality, this temperature is the maximum temperature in the flare region, and corresponds to the superhot component of flares (\cite{ShiYoko2002}). The observed flare temperature is measured when the flare emission measure becomes maximum as a result of evaporation flow. At this time, the flare temperature is a bit smaller and comparable to $10^7$ K. 
 
We can estimate the density of the evaporation flow $n_{\rm ev}$ (cm$^{-3}$) from the balance between enthalpy flux carried by the evaporation flow and the electron beam energy flux $F_e$

\begin{equation}
5 n_{\rm ev} k T C_s  =  F_e \, ,
\end{equation}
where $C_s$ (cm s$^{-1}$) is the sound speed of the heated flare plasma with temperature $T$ estimated to be $C_s=5\times 10^7$ cm s$^{-1}$ ($T/10^7$K)$^{1/2}$. 
Then the density of the evaporation flow becomes

\begin{equation}
n_{\rm ev} = 3 \times 10^{11} {\rm cm}^{-3} \, \bigg(\frac{T}{10^7 \, {\rm K}} \bigg)^{-3/2} \frac{F_e}{10^{11} \, {\rm erg} \, {\rm cm}^{-2} \, {\rm s}^{-1}} \, .
\end{equation}
This is a lower limit of the flare loop plasma density, because the evaporating mass accumulates in the loop so that the flare loop density can increase in time as long as the evaporation flow continues. 
On the other hand, the upper limit of the flare loop density is determined by the balance between the gas pressure of the flare loop plasma and the magnetic pressure which confines the plasma,
$2nkT = B^2 / 8 \pi$.
Hence we obtain

\begin{equation}  
n = 10^{11} {\rm cm}^{-3} \, \bigg(\frac{B}{100 \, {\rm G}} \bigg)^2 \bigg(\frac{T}{10^7 \, {\rm K}} \bigg)^{-1} \, .
\end{equation}
The typical flare loop density in solar flares derived from soft X-rays (for $T$=10$^7$ K) is $10^{11}$ cm$^{-3}$, and so the above order-of-magnitude theory seems consistent with observations assuming $B \simeq 10^2$ G.

Is it possible to have flare loop densities larger than $10^{12}$ cm$^{-3}$, reaching 10$^{13}$ cm$^{-3}$ or even more ?
If the coronal magnetic field in the reconnection region is 300 G, the electron beam energy flux becomes 3 $\times$ 10$^{12}$ erg cm$^{-2}$ s$^{-1}$. In this case $n_{\rm ev}$ 
reaches 10$^{13}$ cm$^{-3}$.  However, such dense flare plasma cannot be confined by 300 G magnetic loop. Instead, we need 1 kG magnetic loop.  Or, if the flare temperature becomes less than 10$^6$ K, then such high density flare plasma can be confined. 

In the case of M dwarfs, it is well known that huge starspots with a few kG are present in some active stars
(e.g. \cite{KrullVal1996}, \cite{Berdyugina2005}).
Furthermore, the pressure scale height is shorter in M dwarf photosphere and chromosphere so that the transition-region height is lower in M dwarf atmosphere than in the solar atmosphere. As a result, the coronal field strength of M dwarfs can be stronger than that on the Sun even if the photospheric magnetic field distribution and strength of M dwarfs are the same as those of the Sun. Altogether, 1 kG magnetic loop is quite likely on M dwarfs, so that the flare loops could be seen as a white light flare on M dwarfs. 

On the other hand, the momentum balance in a flare loop requires that the momentum of the downward moving chromospheric condensation (con) is balanced by the momentum
of the evaporation flow (ev) in a hot loop above the condensation - this was first proven observationally by \cite{Canfield1987}. Assuming that the hydrogen plasma density
is roughly proportional to the electron density, we can write for the momentum balance

\begin{equation}
n_e^{\rm con} V^{\rm con} = n_e^{\rm ev} V^{\rm ev} \, ,
\end{equation}
where $V$ represents the respective flow velocities. From various solar observations we can roughly estimate the ratio $V^{\rm ev}/V^{\rm con}$ to
be around 10 which gives the ratio of electron densities of the same order of magnitude. Typical chromospheric densities derived from flare ribbon observations and
modeling range between 10$^{13}$ and 10$^{15}$ cm$^{-3}$, where the latter value was obtained for an atmosphere strongly heated by electron beams having fluxes
of the order of 10$^{13}$ erg cm$^{-2}$ s$^{-1}$ (\cite{Kowalski+2015}). This then gives an estimate for electron densities in the hot loops in a range
10$^{12}$ to 10$^{14}$ cm$^{-3}$. This is consistent with the above estimates based on the reconnection scenario.

Finally, using this range of electron densities, we can estimate the radiative cooling times for hot 10$^7$ K loops, to be cooled down to temperatures at which we can
see cool loops. Using Eq.(15) for loop densities 10$^{12}$ to 10$^{14}$ cm$^{-3}$, we get the cooling times of the order of 100 to 1 sec, respectively. This then
means that cool loops can be detected quite early, before the flare maximum, and they will produce a significant portion of the WL emission during the gradual
phase. In the solar case, the loop densities are lower and the cool loops appear in spectral lines later, namely at the onset of the gradual phase (see e.g. \cite{Schmieder+1995}). 

\section{Comments on spectral distribution of WL emission}

The mechanisms of the WL continuum
formation in stellar flares, like free-bound
(Balmer, Paschen), as well as free-free, have been considered by many
other authors (see e.g. \cite{HF92} for older work and \cite{Kowalski2016} for a recent review). 
However, those
authors interpreted the WL continuum spectra, in analogy with solar
flares, as arising from the flare
ribbons, i.e. footpoints of the extended loops. 
In the optical range, the Paschen continuum formed at chromospheric levels, is usually optically thin unless the electron-beam
fluxes are very strong (see F13 model of \cite{Kowalski+2015}). In the latter case the Paschen continuum will saturate to blackbody spectrum
with characteristic chromospheric temperatures. However, blackbody emission may also come from deeper photospheric layers presumably
heated by backwarming, but that is detectable only if the overlying forming region of the Paschen continuum is optically thin. Altogether,
such spectral distributions seem to be consistent with typically observed strong blackbody like continua with best fits in the 8 000 - 12 000 K range
(e.g. \cite{HF92}, \cite{Kowalski+2012}). There is also an issue of the Balmer continuum which we don't detail in this study.

A novel aspect of the present paper is to show under which conditions also the flaring loops, overlying large areas and spanning temperatures from
cool to hot, can contribute to the total continuum radiation of stellar flares. As we already mentioned in the introduction, the WL
emission from solar loops was never detected against the disk and this is because the typical electron densities in solar
flare loops don't significantly exceed 10$^{12}$ cm$^{-3}$. The same may apply for standard flares on cool stars, but the
situation may differ in case of superflares where we expect much larger densities (Section 5). 
On the Sun we see both cool and hot loops simultaneously (namely during
the gradual phase) and thus we have also considered a possible contribution of hot loops which
mainly emit due to free-free mechanism. However, we don't say that this must be dominant. If the cool
loops occupy most of the arcade volume, the WL emission from them will be dominated by the 
free-bound Paschen continuum.

There is an important question how can our models be consistent with WL spectral observations of stellar flares, which are
known for a long time to show the blackbody behavior at temperatures around 10 000 K. 
Here we focus just on the flare loops overlying the whole active region and we compute their WL emission, i.e.
not the emission of narrow ribbons. In reality, however, the total stellar
flux will be a mixture of both ribbon and loop components and thus a more sophisticated analysis of the
observed spectra will be required to understand this behaviour. We don't show the computed spectra in
the present paper because they correspond only to the loop component. But we compare our models with typical 
amplitudes of {\em Kepler} superflares in order to set the limits on the loop detectability in WL. In fact there exist
no spectral observations yet of the {\em Kepler} stars during their superflares
so that a comparison with models is impossible and any discussion would be rather premature. 
In this paper we only say that in analogy with the Sun there must be present arcades of cool as well as
hotter loops, presumably much more extended on superflare stars than on the Sun (much larger starspots)
and this is generally accepted by the community. Then if such loops may reach electron densities
around $10^{13}$ or higher (we estimate this based on the reconnection models in Sec. 5), such
loops will substantially contribute to WL continuum emission of {\em Kepler}
superflares. This contribution has to be added to emission arising from the flare ribbons, taking into
account proper (but still largely unknown) filling factors for both types of structures and only then the
model spectra can be compared with future spectral data on superflares. 

Although no superflare from {\em Kepler} sample was yet observed spectroscopically, there is one detection
of very strong flare (called megaflare) on dMe star YZ CMi (\cite{Kowalski+2010}, \cite{Kowalski+2012}). 
The flare spectra show typical blackbody continuum with temperatures around 10 000 K.
However, the flare is much more complex and in fact it shows also strong Balmer (and perhaps
Paschen in the visible) component which, as the authors claim, is
spatially much more extended. Since the spectra were taken during the gradual phase
of this megaflare, one would expect large areas covered by loops, both cool and hot. That might
be consistent with our model which predicts the Paschen (and Balmer) continuum for cool loops
rather than blackbody continuum, unless the Paschen continuum saturates to blackbody at very high loop densities
(see our figures for $T$=10 000 K).

\section{Discussion and Conclusions}

As follows from our modeling, the WL amplitudes due to flare loops will critically depend on electron densities and relative flare areas (filling factors). In the preceding
section we have shown that required electron densities higher than 10$^{12}$ - 10$^{13}$ cm$^{-3}$ are quite expectable as the result of strong evaporative
processes during superflares, although on the Sun such densities are not common. However, relatively large flare areas we used in Figs. 1-4 deserve some discussion. 
From high-resolution observations of solar flares (e.g. \cite{Jing2016}) we clearly see the ribbons as long but narrow features. Such ribbons appear
in WL even less extensive. On the other hand, the total area typically covered by the system of flare loops is much larger.
Contrary to solar flares, in all current analyses of stellar superflares it is assumed that the chromospheric flare condensation is optically thick in the Paschen 
continuum with a representative temperature 10$^4$ K which leads to the blackbody WL spectrum 
(\cite{Shibayama+2013}, \cite{KatLiv2015}, \cite{Kowalski+2015}).
Based on that, one gets relatively small areas which are typically between 0.01 to 1 \% of the whole stellar
surface (Maehara, private communication). For solar type stars with an amplitude 0.01 the area of such
ribbons is only 0.2 \%. However, the same amplitude can be reached by assuming that flare loops cover much larger area, like 10 - 20 \% of the
stellar surface. This is also consistent with areas covered by expected large starspots (\cite{Berdyugina2005}, \cite{Aulanier+2013}, \cite{Maehara+2017}). 
In the case of Sun we see that WL ribbons
are much smaller compared to spot areas (or active-region areas in general), \v{S}vanda et al. (private communication) found a factor around 10 using set
of WL flare and sunspot observations by {\em SDO}/HMI. However, the areas covered by flare-loop arcades are much larger and comparable to size of active regions,
this is well documented on many images from {\em SDO}/AIA.
For dMe stars modeled in Fig. 3 and 4, the required electron densities are smaller, closer to solar ones, or the flare areas needed to reach the
observed amplitudes can be smaller than for solar-like stars - this is due to a higher contrast for stars having low $T_{\rm eff}$.
We thus conclude that the WL emission of superflares can be due to a mixture of WL ribbons and flare loops which are also strongly emitting
in WL. Actual ratio will then depend on the assumed temperatures of ribbons and loops, on the electron density in the loops and on the areas covered by ribbons and
loops. During evolution of the whole flare structure where
the loops grow up in time and can occupy large areas, the WL emission from stellar flare loops can be dominant. On the other hand, we know from solar
observations that the flare ribbons become less
bright during the gradual phase (and mostly invisible in WL) and the whole active region is typically covered by a large system of loops.

In summary, we propose in this paper that extended cool, as well as hotter, loops overlying the
whole active region can significantly contribute to the total flux during flares. This is
based on a close solar analogy with the so called 'post-flare loops'. On the Sun the two
components - ribbons and loops - are spatially well visible in spectral lines and we can study them separately. On the
stars, however, they mix together and thus both may contribute to the total stellar flux in the
optical range. Such scenario is quite novel and certainly deserves further verification and development.

\acknowledgments
{We acknowledge useful discussions with H. Maehara, Y. Notsu, K. Namekata and M. B\'{a}rta and appreciate useful comments of the anonymous referee. 
This work was partially supported by the grant No. 16-18495S of the Czech Funding Agency and by ASI ASCR project RVO:67985815. 
PH also acknowledges the support by the International Research Unit of Advanced Future Studies at Kyoto University during his stay in Japan.}

%\bibliographystyle{aasjournal}
%\bibliography{heinzel}

\end{document}